# Cloud-Enabled Virtual Prototypes

Bridging Local and Remote Simulation

Tim Kraus, Robert Bosch GmbH, Renningen, Germany (*tim.kraus@bosch.com*)

Axel Sauer, Robert Bosch GmbH, Renningen, Germany (*axel.sauer@bosch.com*)

Ingo Feldner, Robert Bosch GmbH, Renningen, Germany (*ingo.feldner@bosch.com*)

*Abstract—* **The rapid evolution of embedded systems, along with the growing variety and complexity of AI algorithms, necessitates a powerful hardware/software co-design methodology based on virtual prototyping technologies. The market offers a diverse range of simulation solutions, each with its unique technological approach and therefore strengths and weaknesses. Additionally, with the increasing availability of remote on-demand computing resources and their adaptation throughout the industry, the choice of the host infrastructure for execution opens even more new possibilities for operational strategies. This work explores the dichotomy between local and cloud-based simulation environments, focusing on the trade-offs between scalability and privacy. We discuss how the setup of the compute infrastructure impacts the performance of the execution and security of data involved in the process. Furthermore, we highlight the development workflow associated with embedded AI and the critical role of efficient simulations in optimizing these algorithms. With the proposed solution, we aim to sustainably improve trust in remote simulations and facilitate the adoption of virtual prototyping practices.**

## I. Introduction

The advent of on-demand computing services has revolutionized how organizations approach their working methods and IT infrastructure. By providing scalable resources, cloud platforms enable businesses to deploy computation tasks flexibly and access powerful processing hardware. This shift has significant implications also for simulation methodologies, particularly in the context of embedded systems development, which traditionally relied on on-site setups derived from working methods that utilized physical hardware. As trust in simulation-based methods for embedded systems has increased, development based on Virtual Prototypes (VPs) has gained traction, gradually replacing physical hardware prototypes while still relying on local setups. To improve set up times and execution performance, utilizing remote compute resources is the next logical step. This transition catches up to the working methods in other domains, like the area of AI algorithms, where developers have ever embraced cloud computing more naturally, in particular for model training.

From a technical perspective, the initial setup of simulation frameworks, which are typically intended for local desktop use, poses challenges when transitioning to a cloud environment. Legally, ensuring the security of protected intellectual property (IP) involved in the simulation - whether in the form of models or the data being processed - becomes critical. Commercially, if compute performance is required for extended periods, investing in local resources may prove more economically efficient than relying solely on cloud services. These aspects further emphasize the need for organizations to evaluate their simulation strategies in the light of available local and remote resources.

In this paper, we extend SUNRISE [1], the *scalable unified RESTful infrastructure for system evaluation*, a framework that simplifies access to virtual prototyping methodologies. By exploring variations of infrastructure back-ends, we analyze the trade-offs between local and cloud-based simulation and we provide insights to enhance the effectiveness and trustworthiness of simulation practices in embedded systems development.

## II. Related Work

The growing importance of simulations, particularly virtual prototypes of embedded systems, is consistently reported by parties of the industry [2]. This trend is increasingly driven by extended usage beyond the early phases of semiconductor development to the post-silicon phase [3], as in the context of digital twins and in combination with physical models. While the shift to cloud-based execution of simulations in hardware/software co-design was predicted early on [4], market-ready solutions have lagged behind. In contrast, the domain of physical simulation, where compute intensive computer-aided engineering (CAE) poses significant challenges, has seen a quicker transition to remote resources [5]. Hybrid approaches have also been proposed, using local resources for simpler tasks while leveraging the cloud for peak requirements, thereby highlighting the need for flexible deployment concepts [6]. With the new diverse infrastructure setups, the deployment of workloads has become a critical area of research. The execution of simulation tasks in heterogeneous environments, with containers or virtual machines and communication between these components are actively discussed in academia [7].

In our work, we deploy simulations as monolithic workloads to hybrid compute back-ends. This must not be confused with the *Hybrid Simulation* scenario, which has been the subject of various discussions in recent years. In contrast to our approach, these solutions split simulations either by running models on different hosts or connecting multiple simulators [8]. The primary goal of these methods is to speed up simulations by improving execution of a single simulation platform, e.g. by combining parts of a classic VP on a desktop computer with emulation on FPGAs for other parts [9] [10]. The *Accellera Federated Simulation Standard Working Group (FSSWG)* [11] focuses on developing methods for connecting simulations and models across different standards, which requires APIs for time synchronization, communication, and data exchange.

For the technical implementation of our workload deployment on cloud resources, we leverage well-established solutions. Our work can be hosted on a simple container runtime or it can be integrated into larger-scale environments that utilize orchestration platforms like Kubernetes [12]. The setup process can be simplified through an *Infrastructure-as-Code* approach, such as Pulumi [13]. Additionally, for running containerized tasks within a host-agnostic workflow model, Argo Workflows is a powerful approach [14] that addresses similar challenges.

III. THE SUNRISE FRAMEWORK

The SUNRISE framework is designed to simplify access to simulation technologies by introducing abstractions that enhance user-friendliness and flexibility. To achieve a high degree of reusability, definitions are made agnostic to the type of simulation workload executed and independent of the underlying compute infrastructure. Therefore, the framework uses a structured approach to represent a generic simulation workflow, encompassing the essential steps: *configure, build, run and analyze*.

A. Overview

The foundation of SUNRISE is the employment of defined APIs (Application Programming Interfaces). The REST interface *Evaluation API (EvalAPI)* is the entry point for virtual prototype users. Simulation models are integrated through the *System API (SysAPI)*, which is especially focused in this work. Figure 1 gives an overview of the components of the SUNRISE framework.

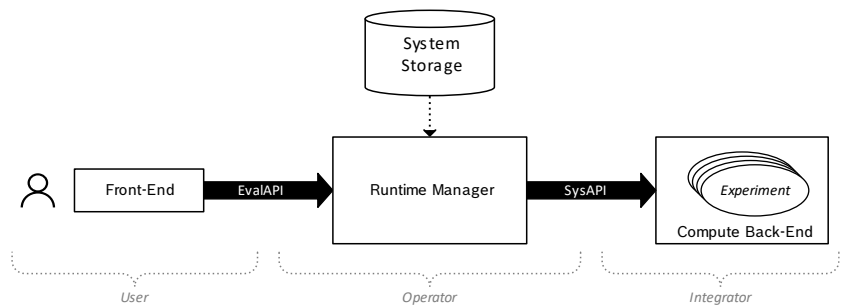

Figure 1: SUNRISE framework overview

Simulation platforms, including all required tools and dependencies, are bundled as so-called *Systems* and stored in the *System Storage*, which can be seen as a library or even a marketplace. The *Runtime Manager* is the central engine that keeps track of the available simulation platforms in the system storage and brings them to execution on a *Compute Back-End* through the SysAPI. The workload execution is logically realized in the form of *Experiments* with an underlying state-machine for each, according to the generic concept of the simulation workflow. For the users of the framework, the Runtime Manager provides the EvalAPI, which can be connected to custom *Front-Ends*. The framework is designed as a hosted service platform, allowing multiple users to utilize the infrastructure for different systems and run numerous experiments in parallel.

*B. Roles*

The SUNRISE methodology distinctly separates the roles of three perspectives: the simulation *user*, the system *integrator* and the infrastructure *operator*. Separating these responsibilities allows each specialist to focus on their specific domain, thereby enhancing overall efficiency in the virtual prototyping ecosystem.

The *user* implements application-specific front-ends against the EvalAPI with no need to know about the technical details related to the setup of the simulation model, tools and environment. This significantly reduces the effort and time overhead for running simulations with a virtual prototype. The realization of a front-end can be rather generic, providing options to modify any configuration parameter available in the system and run through the execution workflow, or it can be designed specifically for a use case by only accessing a subset of parameters and visualizing the relevant experiment results and metrics. Separate solutions for front-ends can be used in parallel for multiple users with different scopes [1].

The system *integrator* is responsible for implementing the SysAPI to make simulations accessible and executable in a defined manner that is fully independent of the application it is used for. This role involves structuring the models in the specified way and setting up a container image with all runtime dependencies, allowing the system to be seamlessly integrated into the SUNRISE framework. As the interface is standardized and common, there is no need for the integrator to train users on specific simulation technologies.

The overall hosting of the framework is managed by an o*perator*, who is responsible for enabling containerized workload deployment and ensuring that the necessary computational resources are available. In this perspective, neither the details of the simulation tools nor the specific use case are relevant, as all components are encapsulated within containers.

*C. System Integration*

The SysAPI plays the key role in extending the SUNRISE framework towards enabling execution of systems on both on-premise and remote back-ends. A central paradigm is the independence of the API from the specific implementation of the functionality inside the system, to support a wide range of virtual prototyping solutions available on the market and even use cases beyond VPs in other domains. Note that the SysAPI for system integration is independent of the EvalAPI for the front-end, which is hosted by the Runtime Manager. It can also be used standalone and locally by the user to run simulations, but this is not within the scope of this work.

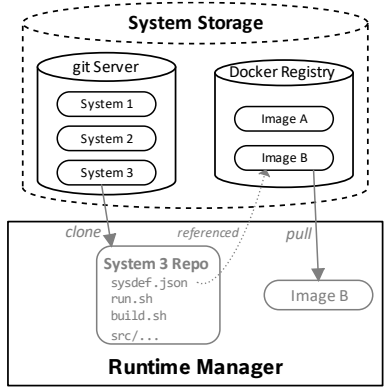
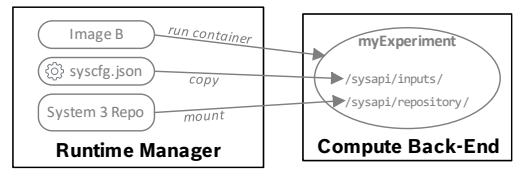

Figure 2: System Components and Integration        Figure 3: Action Execution (Build or Run)

The SysAPI definition comprises a sequence of actions and a set of definitions for files. The API concept is interpreted in a wider sense than just a function interface or communication protocol. The specification of the API includes to the definition of the interface of systems and their functional encapsulation. Structurally, a system in SUNRISE consists of two components (see Figure 2): A *git repository* that holds the *System Definition (SysDef)* JSON file and the components which are provided in source code and compiled during the build step of the workflow, and a *Docker image* that provides the runtime environment of the system, such as the simulation tool along with its dependencies and linked pre-compiled models.

The SysDef file, as shown in Listing 1, contains the name and version to identify the system and the link to the runtime image on a Docker registry. The strings in `build_command` and `run_command` define the action that will be called inside the container to build and run the simulation. The remaining entries define user-configurable parameters that are active for the build and run actions, along with default values for each. Finally, the result objects that are produced by the system during the run action are declared.

To keep track of the systems that are available, it is sufficient for the Runtime Manager to maintain a list of links to repositories, from each of which it can retrieve all information about the system from the included SysDef file. To prepare the workflow execution, when the user requests a new experiment for a system, such as 'System 3', the Runtime Manager clones the repository and pulls the Docker image referenced in the SysDef file. Then for the experiment, a Docker volume is created to transfer data into the Docker container and keep result files afterwards. The following directories are defined for the SysAPI:

- 'repository/' contains the system repository that was cloned from the git server
- 'inputs/' is used for additional data that is provided to the system, like file parameters ('app' in Listing 1)
- 'outputs/' recommended directory for the system to provide data back to the user

```json
{
  "name": "System 3",
  "version": "1.2",
  "docker_image": "my_registry.com/image-b:demo",
  "build_command": "python build.py",
  "run_command": "source run.sh",
  "build_parameters": {
    "compile_args": "-O3 -Wall"
  },
  "run_parameters": {
    "run_time_ms": 1000,
    "app": {
      "default_value": "demo_sw/demo_app",
      "is_file": true
    },
    "simulator_args": "--verbose"
  },
  "results": {
    "signal_trace": {
      "path": "vp/output/sim_trace.vcd",
      "type": "vcd"
    }
  }
}
```

Listing 1: System Definition JSON (SysDef)

```json
{
  "system": {
    "name": "System 3",
    "version": "1.2"
  },
  "build_parameters": {
    "compile_args": "-Os"
  },
  "run_parameters": {
    "run_time_ms": 20,
    "app": "/sysapi/inputs/myApp.elf"
  }
}
```

Listing 2: System Configuration JSON (SysCfg)

The execution of the build and the run action for the system is performed in the same way, as shown in Figure 3. Derived from the SysDef, a *System Configuration (SysCfg)* JSON file is created. As shown in Listing 2, it holds user-modified configuration parameters to be used during the build and run actions. For parameters not present in the SysCfg, the default value from the SysDef will be used. The Runtime Manager copies the SysCfg file into the experiment volume under the 'inputs/' directory.

After these preparation steps, the Docker container can be started to perform an action. The working directory is set to the repository directory in the mounted volume ('/sysapi/repository') and the action call (`build_command`/`run_command`) from the SysDef file is appended as command to `docker run` call, followed by the absolute path of the SysCfg JSON file. The complete command line to perform a simulation run like the Runtime Manager does, looks as follows:

```
docker run --rm  -v <experiment-volume>:/sysapi \
                 -w /sysapi/repository \
                 my_registry.com/image-b:demo \
                 source run.sh /sysapi/inputs/syscfg.json
```

If any result files which are produced during the simulation run shall be available to the user for analysis, they must reside in the mounted volume, as after executing a containerized action, the container itself is removed.

The described system integration and execution is realized fully containerized, allowing it to be performed on almost any host system. The storage of systems relies on common, open-source friendly and industry standard technologies like git and Docker. All of this enables the flexible deployment of the simulation workload, as described in the following chapter.

IV. NOVEL DEPLOYMENT CONCEPTS

Traditionally, the deployment of simulation workloads has been characterized by monolithic environments that are closely tied to specific infrastructure requirements, tools, and data formats. This approach is limited in terms of scalability and flexibility, making it challenging to adapt to evolving computational needs.

To address these limitations, we propose to extend the SUNRISE framework for better scalability by enabling the invocation of simulations across a variety of infrastructure solutions. This includes leveraging on-premise systems, utilizing external cloud providers or adopting a hybrid approach that combines both. Each category will be elaborated upon in the subsequent sections, highlighting how they can be effectively utilized to split workloads between local and remote resources. In Figure 4 an overview of the new deployment concepts with the SUNRISE framework is shown.

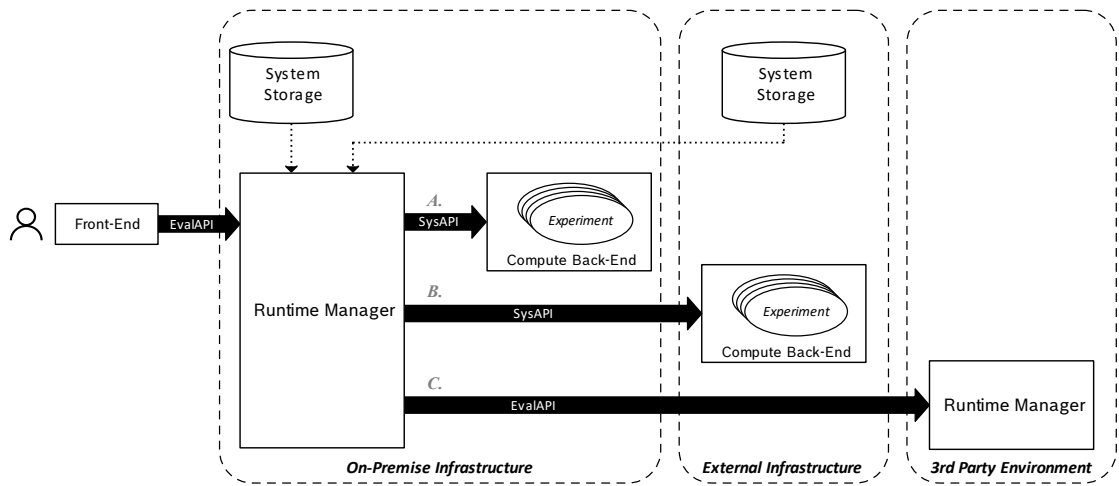

Figure 4: SUNRISE Infrastructure Deployment Concepts

*A. On-Premise Resources*

On-premise resources provide users with full control over local host systems, albeit at the price of continuous operational expenses. This option ensures absolute privacy for confidential data belonging to users but assumes that models developed by suppliers are delivered to the user and can be run in his environment. The initial implementation of the SUNRISE framework was based on this scenario, necessitating the hosting and maintenance of the infrastructure, as well as the acquisition of appropriate server machines and management of access rights.

*B. External Resources*

External resources, such as the major large scale cloud computing platforms, promise an increase in compute power. Nevertheless, single-core performance, which is crucial for e.g. for SystemC based virtual prototypes, is often even lower compared to on-premise workstation computers. The performance benefits in this context are primarily realized through parallelization. The SUNRISE framework utilizes a cloud provider back-end through a vendor agnostic wrapper based on an *Infrastructure-as-Code* approach [13].

A key benefit of on-demand services is that costs will only incur during the active simulation, as all necessary resources are provisioned exclusively for its duration. Upon completion, these resources are decommissioned, while the simulation data is retained, enabling subsequent repetitions or variations of the simulation with different parameters. The dedicated resource allocation for each experiment facilitates significant performance improvements when multiple simulations are executed in parallel. While SUNRISE incorporates cloud resources, it is not designed as an orchestration system itself. Instead, our focus is on developing interfaces, workflows, and data formats in the context of VP simulation.

*C. Cascaded EvalAPI*

The cascaded EvalAPI represents an innovative method for integrating systems within the SUNRISE framework. The core idea is to use the front-end EvalAPI also in the back-end to connect another Runtime Manager, as an alternative solution to providing a system based on the SysAPI. It is important to highlight that the EvalAPI can be utilized as specified without modifications. Used as back-end API, it allows effectively abstracting and concealing the systems provided by vendors. In contrast to integrating systems with the SysAPI, this approach ensures complete isolation of the back-end implementation and the data involved from the Runtime Manager, guaranteeing that IP vendors maintain full privacy and control over their provided systems.

In summary, the cascaded EvalAPI offers significant benefits for system integrator, but also introduces challenges:
- Vendor Control: All system artifacts remain with the system integrator, eliminating the need to share or expose them, which enhances security and confidentiality.
- Resource Utilization: The cascaded EvalAPI allows vendors to use their chosen resources for running and executing systems (e.g. internal servers), providing flexibility in resource management.
- Implementation Flexibility: System vendors can select their own internal operation concepts, allowing the use of alternatives to Docker images and Git repositories.
- Custom Hardware Selection: Vendors can choose custom or specialized hardware to execute their systems, such as FPGAs, prototyping boards or emulators, enhancing performance and adaptability.
- Hosting Responsibility: This approach necessitates hosting a dedicated REST API server, which requires competencies to provide and secure an internet-facing API

*D. Hybrid Compute Infrastructure*

The previously presented concepts discussed the static deployment of experiments to a specific type of computation back-end. A major improvement for SUNRISE in this work, compared to initial concepts [1], is the support of a hybrid compute infrastructure. Experiments can be assigned to one of multiple back-end options based on user selection or system requirements at the time an experiment is created. The EvalAPI at the front-end is used the same way for any back-end, ensuring that users can interact with the systems seamlessly, regardless of the underlying infrastructure.

Moreover, the hybrid compute infrastructure supports concurrency through the Runtime Manager that can handle multiple infrastructures simultaneously. This capability allows for the parallel execution of experiments across multiple back-ends, optimizing resource utilization and reducing overall processing time. The ability to choose a back-end not only enhances operational flexibility but also addresses privacy requirements for both users and system integrators. By allowing model execution to shift between internal and external resources on a per-experiment basis, the SUNRISE framework ensures that sensitive data and models are handled in compliance with privacy regulations whereas uncritical experiments can benefit from a free choice of the compute back-end.

V. CASE-STUDY: OPTIMIZING AI WORKLOADS FOR EMBEDDED HARDWARE

Bringing inference of AI based algorithms on embedded devices is a challenging task due to the growing complexity and variety of neural networks and microcontroller architectures. This task is addressed by AI deployment toolchains [15], that compile a trained neural network to source-code that can be included in software projects for embedded systems, typically including optimizations to enhance execution performance and minimize memory footprint.

In this case study, we use the SUNRISE framework to improve an AI deployment toolchain for specific Instruction Set Architectures (ISAs), including custom ISA extensions and AI accelerator IP by conducting tests on virtual prototype simulations. We utilize the hybrid compute infrastructure concepts for workload distribution across various back-ends to perform extensive measurements of AI model layers for several architectures using an external cloud platform which maximizes scalability and enables rapid prototyping. To address privacy concerns, a dual simulation strategy is implemented. The simulation with the full AI model is executed on an on-premise workstation to evaluate performance in a controlled environment, while specific components are tested on novel AI accelerators within the secure infrastructure of the IP vendor using the cascaded EvalAPI approach. Figure 5 shows the utilization of the hybrid compute infrastructure concept.

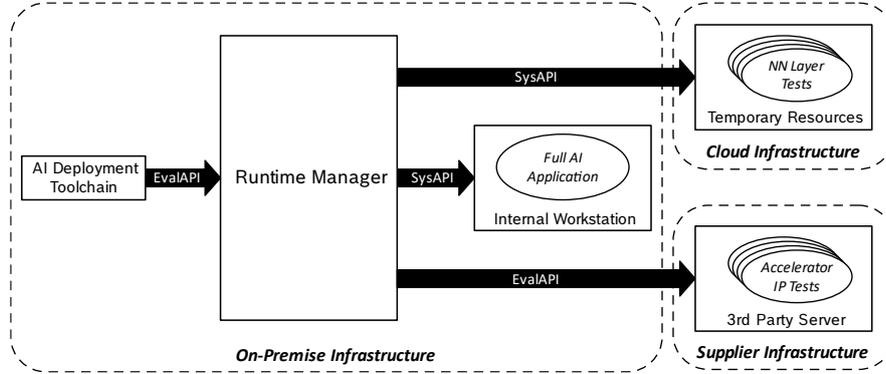

Figure 5: Application with an Embedded AI Toolchain

*A. Approach*

The AI algorithms consist of various neural network layers, such as fully-connected and convolutional layers, which are initially benchmarked separately. Performance optimizations are conducted layer-by-layer for specific embedded processor cores, primarily targeting RISC-V architectures including specialized packed-SIMD instructions to accelerate neural network operations. Given the large size of the design space that is explored, the measurements of individual layers on the different systems are performed using cloud infrastructure. Subsequently, the optimizations for each layer are integrated and tested collectively across multiple neural network models in an on-premise infrastructure. Concurrently, evaluations are carried out on a dedicated neural network accelerator IP, which can compute individual parts of the neural network or specific layers. This testing is conducted within the infrastructure of the accelerator IP vendor, utilizing a cascaded API approach. The accelerator features vector extensions that are implemented close to the core, as well as an external accelerator attached to the SoC interconnect.

*B. Results*

To assess the overall success of the case study utilizing the SUNRISE framework with its hybrid compute infrastructure concept, a Key Performance Indicator (KPI) for productivity is derived which addresses the total simulation time across all optimization runs. The times are recorded in a full test setup with $n_{layer} = 18$ network layers, $n_{opt} = 6$ optimization permutations per layer and $n_{arch} = 9$ target hardware architectures. This setup results in a total of $n_{layer} \cdot n_{opt} \cdot n_{arch} = 972$ simulations for each variant of the AI toolchain that is tested. Each simulation takes approximately two minutes in average, leading to a total compute time of around 32 hours for a single-core execution. To enhance efficiency, we explored various parallelization options across different compute environments:

- Personal Computer: Capable of running simulations with a factor of x4 parallelization.
- Workstation Computer: Offers a higher parallelization factor of x8.
- Cloud Service: Theoretically allows for unbounded parallelization. We used the maximum parallelization of all simulations resulting in a factor of x972, while being aware of the inherent limitations imposed by cloud service providers regarding resource allocation and scalability.

Table 1 summarizes the total execution times across different environments, illustrating the significant impact of parallelization using cloud resources. A single simulation consumes around 50% more time compared to the other compute envirogrtents due to resource provisioning and slower single-core performance. The results show that the hybrid compute infrastructure of SUNRISE significantly speeds up and simplifies the optimizations of the deployment of AI based algorithms on microcontroller architectures. Regarding the performance of the deployed AI application on the target hardware, the deployed neural network has been significantly accelerated for the new RISC-V targets and external accelerators, with notable improvements in memory optimization.

| Compute Environment | Execution Time |
|---|---|
| Personal Computer | 16 hours |
| Workstation Computer | 8 hours |
| Cloud Service | 3 minutes |

Table 1: Execution times for different compute back-ends

## VI. CONCLUSION

In this paper, we have presented a hybrid compute infrastructure approach based on the SUNRISE framework, which enables the flexible deployment of simulation workloads across a variety of compute hosts. Based on the requirements of users and model suppliers, local and remote back-ends were selected. Our exploration of the trade-offs between scalability, performance and privacy has highlighted the importance of flexible infrastructure options in the field of virtual prototyping. The case study on optimizing AI workloads for embedded hardware illustrated the practical benefits of our approach, showcasing significant reductions in both bring-up time and overall simulation execution time. Future work will focus on refining these deployment concepts and exploring the utilization of specialized hardware, such as physical prototyping boards and FPGA-as-a-Service (FaaS), to expand the range of applications. Additionally, we will explore the integration of encrypted computing solutions to enhance data protection and address evolving privacy concerns.

## ACKNOWLEDGMENT

This work is partially funded by the European Union's HORIZON KDT grant agreement No. 101095947 (TRISTAN - Together for RISC-V Technology and Applications).